\documentclass{article}
\usepackage{spconf,amsmath,graphicx}
\usepackage{microtype}
\usepackage{booktabs} 
\usepackage{multirow}
\usepackage{comment}
\usepackage{array}
\usepackage{xcolor}
\usepackage{colortbl}
\usepackage{caption}
\usepackage{float} 
\usepackage{enumitem}
\setlist{nosep, leftmargin=14pt}
\usepackage{multicol}
\usepackage[hyphens]{url} 

\title{A Deep Learning Framework for Thyroid Nodule Segmentation and Malignancy Classification from Ultrasound Images}
\name{Omar Abdelrazik$^{1}$, Mohamed Elsayed$^{2}$, Noorul Wahab$^{1}$, Nasir Rajpoot$^{1}$, Adam Shephard$^{1}$}

\address{$^{1}$Tissue Image Analytics Centre, Department of Computer Science, University of Warwick, UK \\
    $^{2}$Royal Stoke University Hospital, University Hospitals of North Midlands, UK}

\begin{document}
%
\maketitle
\begin{abstract}
Ultrasound-based risk stratification of thyroid nodules is a critical clinical task, but it suffers from high inter-observer variability. While many deep learning (DL) models function as ``black boxes," we propose a fully automated, two-stage framework for interpretable malignancy prediction. Our method achieves interpretability by forcing the model to focus only on clinically relevant regions. First, a TransUNet model automatically segments the thyroid nodule. The resulting mask is then used to create a region of interest around the nodule, and this localised image is fed directly into a ResNet-18 classifier. We evaluated our framework using 5-fold cross-validation on a clinical dataset of 349 images, where it achieved a high F1-score of 0.852 for predicting malignancy. To validate its performance, we compared it against a strong baseline using a Random Forest classifier with hand-crafted morphological features, which achieved an F1-score of 0.829. The superior performance of our DL framework suggests that the implicit visual features learned from the localised nodule are more predictive than explicit shape features alone. This is the first fully automated end-to-end pipeline for both detecting thyroid nodules on ultrasound images and predicting their malignancy.
\end{abstract}

\section{INTRODUCTION}
\label{sec:intro}

Thyroid nodules are highly prevalent, with detection rates increasing due to the widespread use of high-resolution ultrasound. While the vast majority are benign, accurately distinguishing them from the small percentage of malignant nodules is crucial for patient management. To standardise this process, clinical guidelines such as the British Thyroid Association's (BTA) U-classification \cite{bta2014guidelines} and the American College of Radiology's (ACR) TI-RADS \cite{tessler2017acr} have been established.

A significant challenge with these systems, however, is that their application relies on subjective interpretation of nodule features, leading to considerable inter- and intra-observer variability among radiologists \cite{ataide2020machine}. This has motivated the development of Computer-Aided Diagnosis (CAD) systems to provide objective and reproducible classifications.

In recent years, many CAD systems have demonstrated high performance. Some use machine learning models trained on ``radiomic" features \cite{vadhiraj2021hybrid}, while others utilise end-to-end deep learning models like convolutional neural networks (CNNs) \cite{xu2025comparison, wang2025thyronetx4}. While these ``black-box" models can achieve high predictive accuracy \cite{ataide2020machine, vadhiraj2021hybrid}, their lack of transparency provides a significant barrier to clinical trust and adoption.

To bridge this gap, we propose a new, fully automated deep learning framework for interpretable thyroid nodule malignancy prediction. Our framework achieves interpretability by forcing the model to learn features \textit{only} from the clinically relevant area. This two-stage pipeline first uses a TransUNet \cite{chen2021transunet} model to accurately segment the nodule. The nodule region of interest (ROI) is then fed directly into a ResNet \cite{he2016deep} classifier. This approach mimics the clinical workflow of ``find and assess" and ensures the model's decision is based only on the nodule's visual features (e.g. texture, echogenicity, and implicit shape).

To validate our proposed framework, we compare it against a strong interpretable baseline: a traditional machine learning (Random Forest) model trained on 15 explicit morphological features (e.g. solidity, eccentricity) extracted from the same segmentation masks \cite{kwak2011tirads}. This allows us to directly compare the predictive power of our implicit, region-based DL framework against an explicit, feature-based ML approach.

The main contributions of this paper are:
\begin{enumerate}
    \item We propose a fully automated, two-stage deep learning framework for thyroid nodule classification that localises the nodule before analysis.
    \item We validate this framework on a clinical dataset, demonstrating high performance for malignancy prediction.
    \item We provide a direct comparison of our region-based (implicit feature) DL model against a strong, feature-based (explicit shape) ML baseline, showing that our proposed framework is superior.
\end{enumerate}

\section{MATERIALS AND METHODS}
\label{sec:methods}

\subsection{The Datasets}
\label{ssec:data}

\begin{figure*}[t]
  \centering
  \includegraphics[width=\textwidth]{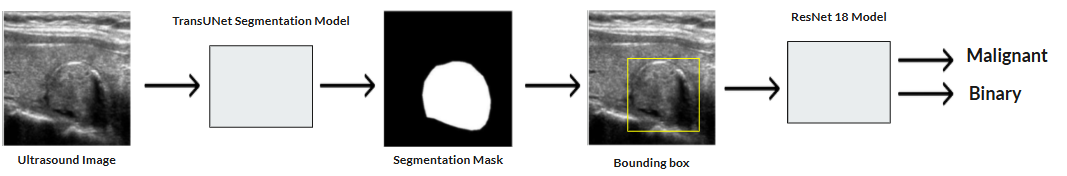}
  \caption{Overview of the two-stage approach for thyroid nodule segmentation and malignancy classification.}
  \label{fig:pipeline}
\end{figure*}

This study employed three ultrasound imaging datasets: DDTI \cite{pedraza2015ddti}, TN3k \cite{gong2021tn3k} and TNUI \cite{ma2022tnui}. The \textbf{DDTI} dataset comprises 480 images from 390 patients, with radiologist annotations for both thyroid nodule segmentation and TI-RADS labels. The \textbf{TN3k} dataset contained 3,493 images, with nodule level segmentations. The \textbf{TNUI} dataset consisted of 1,381 images, with nodule level segmentations.

For the first segmentation task, we trained the model based on the DDTI and TN3k datasets, holding the TNUI dataset back for external testing. For the downstream classification experiments, we utilised the DDTI dataset alone and combined the TI-RADS labels to form a binary classification task of benign (TI-RADS categories 1, 2 and 3) vs malignant (TI-RADS categories 4 and 5). The final dataset for the thyroid classification task consisted of all DDTI samples for which both a valid mask and a corresponding binary TI-RADS label existed, totaling 349 images (61 benign, 288 malignant).

\subsection{Approach}
\label{ssec:overview}

We developed a multi-stage deep learning-based pipeline which first segments thyroid nodules in ultrasound images. We then fed the cropped ROI into a CNN for malignancy classification. An overview of this approach is presented in Fig. \ref{fig:pipeline}.

\subsection{Thyroid Nodule Segmentation}
\label{ssec:segmentation}

The first stage of our pipeline focusing on thyroid nodule segmentation. We primarily employed a TransUNet architecture, which combines Transformer-based global context modelling with the U-Net \cite{ronneberger2015u} architecture, and compared it against a ResUNet baseline \cite{diakogiannis2020resunet}. 

Models were trained on combinations of the DDTI and TN3K datasets, while the TNUI dataset was reserved exclusively for external testing to assess generalisation. For each training configuration, we applied an 80/20 split of the available data into training and validation sets. Training used a combination of dice loss and binary cross-entropy as the loss function. Optimisation was performed using AdamW with a learning rate of $1\times10^{-4}$, and models were trained for 50 epochs with early stopping based on validation performance.

Segmentation performance was evaluated using Dice Score and Intersection over Union (IoU) across internal and hold-out test sets. Table \ref{tab:segmentation_results} summarises results for all train–test combinations, highlighting the superior robustness of TransUNet compared to ResUNet.

\subsection{Thyroid Nodule Classification}
\label{ssec:approach_a}
\subsubsection{Deep Learning Approach}
\label{sssec:deep_learning}

After segmentation, we then extracted regions of interest (ROIs) centred on the thyroid nodules, but expanded with a 10-pixel padding to further include immediately surrounding tissue that may be beneficial to the classification task. This ROI is resized to 224$\times$224, converted to three channels, and normalised using ImageNet statistics.

We employed a ResNet-18 model, pre-trained on ImageNet, for binary classification of the thyroid nodule as benign or malignant. The network was fine-tuned using AdamW with a weighted cross-entropy loss to address class imbalance, assigning higher penalties to misclassified benign cases.

\subsubsection{Morphological Features Approach}
\label{sssec:morphology}

For comparison, we additionally trained a traditional machine learning pipeline using hand-crafted morphological features for classification. The goal of this approach was to quantify nodule morphology. For each thyroid nodule mask (gained from the segmentation task), a vector of 15 distinct features was computed. These features can be categorised as follows:
\begin{itemize}
    \item \textbf{Geometric properties:} Basic measurements including area, perimeter, convex area, and filled area.
    \item \textbf{Irregularity metrics:} Features sensitive to margin irregularity. This included solidity (the ratio of the mask's area to its convex hull area) and form factor (a measure of circularity deviation).
    \item \textbf{Elongation metrics:} Features to quantify nodule shape, such as eccentricity (0 for a circle, 1 for a line) and aspect ratio (major axis length / minor axis length).
    \item \textbf{Complex shape descriptors:} The seven Hu moments, which provide a robust ``fingerprint" of the nodule's shape invariant to translation, scale, and rotation.
\end{itemize}

The extracted 15-feature vector (per nodule) was used to train two widely-used classifiers: a Random Forest (RF) and a multi-layer perceptron (MLP). We first normalised the features using the standard scaler. To handle the significant class imbalance (288 malignant vs 61 benign) we implemented synthetic minority over-sampling technique (SMOTE) \cite{chawla2002smote} to oversample the minority benign class.

This comparison of methods allows us to directly compare the predictive power of engineered, human-understandable features against the learned, implicit features from a deep model focused on the same region.

\subsubsection{Evaluation Methodology}
\label{ssec:evaluation}

Both the deep learning and the morphological feature-based approaches were evaluated using the same stratified 5-fold cross-validation framework to ensure a fair comparison. The dataset was split into 5 folds, ensuring the ratio of benign to malignant samples was preserved in each fold. Performance was assessed by aggregating the predictions from all 5 validation folds. The primary metrics included accuracy and F1-score, recall and precision.

\begin{table}[t]
\centering
\newcolumntype{L}[1]{>{\raggedright\arraybackslash}m{#1}}
\newcolumntype{C}[1]{>{\centering\arraybackslash}m{#1}}
\begin{tabular}{L{1.4cm} L{1.5cm} L{1.3cm} C{1cm} C{1cm}}
\toprule
\textbf{Train Set} & \textbf{Test Set} & \textbf{Model} & \textbf{Dice} & \textbf{IoU} \\
\midrule
DDTI & DDTI (internal) & ResUNet & 0.7800 & 0.6808 \\
DDTI & TN3K & ResUNet & 0.5763 & 0.4668 \\
\midrule
DDTI & DDTI (internal) & TransUNet & 0.8548 & 0.7477 \\
DDTI & TN3K & TransUNet & 0.6300 & 0.5176 \\
DDTI & TNUI & TransUNet & 0.5276 & 0.4141 \\
\midrule
TN3K & TN3K (internal) & TransUNet & 0.8695 & 0.7723 \\
TN3K & DDTI & TransUNet & 0.6757 & 0.5701 \\
TN3K & TNUI & TransUNet & 0.7591 & 0.6608 \\
\midrule
TN3K + DDTI & TN3K + DDTI (internal) & TransUNet & 0.8624 & 0.7639 \\
TN3K + DDTI & TNUI & TransUNet & 0.7935 & 0.6968 \\
\bottomrule
\end{tabular}
\caption{Segmentation performance comparison of ResUNet and TransUNet across different training and testing configurations.}
\label{tab:segmentation_results}
\end{table}

\section{Results}
\label{sec:results}

\subsection{Thyroid Nodule Segmentation Performance}
\label{ssec:seg_results}

We conducted an exhaustive comparison of two segmentation architectures (ResUNet and TransUNet) across multiple training configurations using the DDTI, TN3k and combined datasets, with TNUI reserved for external testing. The full results are presented in Table \ref{tab:segmentation_results}.

The analysis confirmed two key points. Firstly, the TransUNet model consistently yielded superior performance over the ResUNet architecture. Secondly, training on the combined DDTI and TN3k datasets provided the greatest stability and generalisation capability across unseen data.

\begin{figure*}[t]
  \centering
  \includegraphics[width=0.95\textwidth]{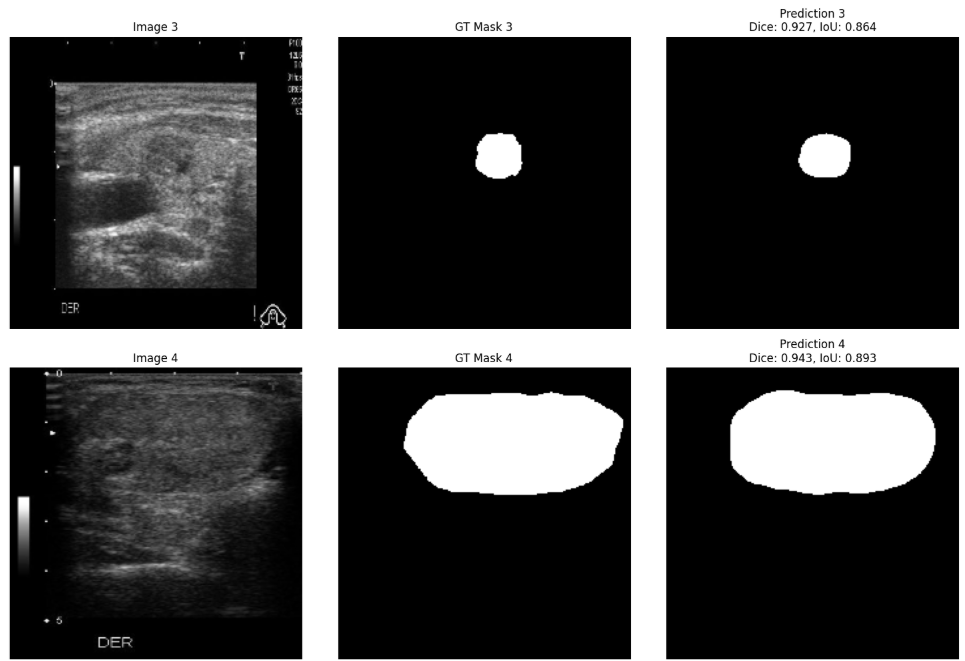}
  \caption{Qualitative segmentation results from the proposed model. Each row shows a different thyroid nodule example with three panels: (left) the original ultrasound image, (middle) the ground truth mask annotated by a radiologist, and (right) the predicted mask generated by the model.}
  \label{fig:seg_results}
\end{figure*}

Based on these findings, we selected the TransUNet model trained on the combined dataset for subsequent classification. When evaluated on the DDTI test set (the cohort used for classification), this model achieved a Dice Score of 0.8624 and an Intersection over Union (IoU) of 0.7639. This high level of accuracy ensured that the masks provided a reliable basis for subsequent ROI extraction. Qualitative examples of segmentation outputs are shown in Fig. \ref{fig:seg_results}.

\subsection{Thyroid Nodule Classification Performance}
\label{ssec:class_results}

We evaluated our two distinct approaches for binary TI-RADS classification (Benign: 1-3 vs. Malignant: 4-5) using the 5-fold cross-validation methodology. The aggregated performance of all three classifiers is directly compared in Table \ref{tab:main_results}. We see that our deep learning-based approach achieves the highest results (Dice = 0.8522), with our morphological feature-based approach also gaining high performance (RF Dice = 0.8294).   

\begin{table}[t] 
\centering

\newcolumntype{L}[1]{>{\raggedright\arraybackslash}m{#1}}
\newcolumntype{C}[1]{>{\centering\arraybackslash}m{#1}}
\begin{tabular}{L{2.5cm} C{1cm} C{1cm} C{1cm} C{1cm}}
\toprule
\textbf{Model} & \textbf{F1} & \textbf{Acc.} & \textbf{Recall} & \textbf{Prec.} \\
\midrule
ResNet-18 & 0.8522 & 0.7822 & 0.8229 & 0.8843 \\
Random Forest (Morph. features) & 0.8294 & 0.7278 & 0.8024 & 0.8556 \\
MLP (Morph. features) & 0.7649 & 0.6619 & 0.6713 & 0.8963 \\
\bottomrule

\end{tabular}
\caption{Aggregated 5-fold cross-validation results for thyroid nodule classification as benign or malignant.}
\label{tab:main_results}
\end{table}

\section{Discussion}
\label{ssec:discussion}

In this work, we aimed to develop a fully automated end-to-end approach for segmenting thyroid nodules on ultrasound images and further classifying them as benign vs malignant. The results demonstrate that both deep learning and classical approaches using interpretable morphological features are highly effective for malignant classification.

The proposed deep learning model achieved the highest overall performance, with a F1-score of 0.852, outperforming the morphological feature based approach. This suggests that while shape is a dominant predictor, ResNet-18 is able to learn additional, \textit{implicit} visual patterns from the cropped nodule's texture and echogenicity that provide a slight performance edge over the \textit{explicit} shape features alone. Further, the model correctly identified 237 of the 288 malignant nodules (recall = 0.823). Thus, the low number of false negatives is clinically promising. The main source of error was in misclassifying benign nodules as malignant (25 false positives), an error that typically leads to further follow-up.

The Random Forest model, using only 15 shape features, achieved a high F1-score of 0.829, showing that morphological features are sufficient on their own for classification. This aligns with clinical guidelines, which weigh features like ``taller-than-wide" and ``irregular margins" heavily \cite{tessler2017acr}.


While our results demonstrate the success of segmentation-driven classification, this study has several limitations. Firstly, the performance of both classification approaches is critically dependent on the accuracy of the nodule segmentation. Any errors in the TransUNet's predicted mask will propagate directly to the downstream classification models.

Secondly, the classification task was performed on the 349 labelled images from the DDTI dataset. Although stratified 5-fold cross-validation was used, the dataset is highly imbalanced (288 malignant vs 61 benign). This imbalance, though in-part addressed with SMOTE and weighted loss, makes it challenging for the models to learn the features of the minority benign class, as seen in Table \ref{tab:main_results}.

Finally, the traditional ML pipeline relied \textit{solely} on morphological features, ignoring texture, echogenicity, and calcifications. In constrast, the ResNet model used \textit{only} the cropped image. A future, more complex approach that fuses explicit shape features with implicit deep features could potentially yield even higher performance.

\section{Conclusion and Future Directions}
\label{sec:conclusion}

This study presented a segmentation-driven framework for thyroid nodule malignancy prediction, combining interpretability with strong predictive performance. Our two-stage deep learning pipeline, leveraging TransUNet for segmentation and ResNet-18 for classification, achieved the highest F1-score (0.852) and demonstrated clinically relevant sensitivity, correctly identifying the majority of malignant cases. The comparison with a Random Forest model trained on explicit morphological features (F1-score = 0.829) confirms that shape remains a dominant predictor, but \textit{implicit} deep learning features provide an additional edge by capturing texture and echogenicity.

Despite these promising results, performance is influenced by segmentation accuracy and dataset imbalance, highlighting the need for larger, balanced datasets. Future work will focus on integrating \textit{explicit} shape descriptors with \textit{implicit} deep features, whilst extending to multi-class TI-RADS prediction, and validating on diverse cohorts. By prioritising interpretability and clinical alignment, this approach represents a step toward trustworthy AI-assisted thyroid nodule assessment.

\section{COMPLIANCE WITH ETHICAL STANDARDS}
\label{sec:ethics}

This research study was conducted retrospectively using open access human subject data. Ethical approval was not required.

\section{ACKNOWLEDGMENTS}
\label{sec:acks}

This work was supported by the Department of Computer Science at the University of Warwick.

\bibliographystyle{unsrt}
\bibliography{references}

\end{document}